# Relieving the Need for Bi-Level Decision-Making for Optimal Retail Pricing via Online Meta-Prediction of Data-Driven Demand Response of HVAC Systems

Ah-Yun Yoon, Youngjin Kim*, Seung-Ill Moon

Department of Electrical Engineering, Pohang University of Science and Technology (POSTECH), 77 Cheongam-Ro, Pohang, Gyeongbuk, South Korea 37673
E-mail: powersys@postech.ac.kr

**Abstract:** Price-based demand response (DR) of heating, ventilating, and air-conditioning (HVAC) systems is a challenging task, requiring comprehensive models to represent the building thermal dynamics and game theoretic interactions among participants. This paper proposes an online learning-based strategy for a distribution system operator (DSO) to determine optimal electricity prices, considering the optimal DR of HVAC systems in commercial buildings. An artificial neural network (ANN) is trained with building energy data and represented using an explicit set of linear and nonlinear equations, without physics-based model parameters. An optimization problem for price-based DR is then formulated using this equation set and repeatedly solved offline, producing data on optimal DR schedules for various conditions of electricity prices and building thermal environments. Another ANN is then trained online to directly predict DR schedules for day-ahead electricity prices, which is referred to as meta-prediction (MP). By replacing the DR optimization problem with the MP-enabled ANN, an optimal electricity pricing strategy can be implemented using a single-level decision-making structure, which is simpler and more practical than a bi-level one. In simulation case studies, the proposed single-level strategy is verified to successfully reflect the game theoretic relations between the DSO and commercial building operators, so that they effectively exploit the operational flexibility of the HVAC systems to make the DR application profitable, while ensuring the grid voltage stability and occupants' thermal comfort.

## 1. Introduction

Heating, ventilating, and air-conditioning (HVAC) systems represent more than 35% of the electricity usage of commercial buildings, accounting for approximately 13% of the total electricity consumption in the United States in 2017 [1]. Meanwhile, the controllability of HVAC systems has been improved, mainly due to the development of variable speed drives (VSDs) and building energy management systems (BEMSs) [2]. Therefore, various demand response (DR) programs have been studied, for example, in [3] and [4], where the power inputs of HVAC systems are adjusted in response to variations in electricity price.

### 1.1. Modeling of HVAC systems and building thermal dynamics

To implement DR programs, many researchers have modelled the thermal responses of building rooms to HVAC system operations. In several studies (e.g., [3] and [4]), HVAC units were modelled as point-loads, and indoor temperatures were estimated using simplified equivalent thermal parameter (ETP) models. However, it was reported in [5] that such ETP models would not be appropriate for representing the thermal dynamics of a commercial building room. In [6] and [7], HVAC systems were modelled in more detail based on the dynamic characteristics of physical components, such as heat exchangers and compressors. Indoor temperature variations were also estimated using comprehensive thermal response models [6], [7], considering the size of the building structure and construction materials. However, commercial building operators (CBOs) and end-users are seldom informed of the model parameters of HVAC units and building structures: e.g., refrigerant pressures, time-varying coefficients of performance, and thermal resistance and capacitance of concrete floors and walls. This makes it difficult to implement DR programs in practice. Experimental data-driven approaches have been used to overcome this difficulty [8], [9], where regression algorithms were adopted to develop the polynomial equation models. The experimental models are relatively accurate because HVAC systems can be tested over a wide operating range. However, it is time consuming and costly to repeat these experiments for various types and sizes of HVAC systems and commercial buildings [8]. It is also difficult to test HVAC units that are already in daily service in buildings.

The aforementioned issues can be resolved by using big data and machine learning (ML) algorithms. The HVAC system operation and building thermal conditions (e.g., the weather, indoor temperatures, and occupants' activities) can be better monitored and analyzed when Internet-of-Things sensors are installed and connected to a BEMS. In particular, artificial neural networks (ANNs) can be trained using sensor and third-party data, accumulated in the BEMS database, to reflect the characteristics of the HVAC system operations and corresponding indoor temperature variations. ANN-based models can readily be implemented for various types and sizes of HVAC systems and buildings that are currently in daily service [10], compared to the experimental models. Data volume and variability affect the performance of ANN-based modeling approach, implying the risk that ANNs are over-fitted to limited sets of training data particularly when the approach is applied to new buildings. Recently, online learning has been discussed to mitigate this risk [10], gradually improving modeling performance with additional data collection.

### 1.2. Optimal electricity pricing considering price-based DR

Using such HVAC system models, optimal price-based DR strategies have been extensively studied to reduce the electricity costs of buildings [11], [12]. Furthermore, optimal strategies for electricity pricing have been widely discussed to assist distribution system operators (DSOs) in inducing HVAC load shifts, and hence achieving more cost-effective and reliable operations of the distribution grid: e.g., grid voltage support [11] and peak load shaving [12]. To integrate both DR and pricing strategies, it is essential to characterize the responses of HVAC loads to variations in electricity prices. For example, in [13] and [14], analytic functions were used to model the response characteristics. The price sensitivity coefficients were presented in [15] and [16]. These methods are simple but inaccurate, not reflecting the optimal, inter-time responses of the HVAC loads. In practice, the optimal power input of an HVAC system at time *t* is affected by not only the electricity price at *t*, but also the power inputs and prices at $\tau \neq t$, when the indoor temperatures should be maintained within pre-determined limits.

To avoid this limitation, a bi-level decision-making structure has been adopted using Stakelberg (or leader-and-followers) game theory [17], based on the hierarchical relations between the DSO and CBOs. Given the optimal electricity price, the optimal, inter-time response of the HVAC load can be determined directly in the low-level optimization problem, which affects the optimal pricing in the upper-level problem at the same time. To solve the optimization problems simultaneously, the DSO needs to reformulate the bi-level structure to the equivalent single-level one, for example, by using the Karush–Kuhn–Tucker (KKT) conditions [17]. The reformulation is mathematically onerous due to a number of auxiliary equations and decision variables required to equivalently express the complementary slackness conditions [17]. The reformulated problem is often computationally expensive to solve and hence inapplicable to a large-scale distribution grid including a number of HVAC units [18]. Moreover, the DSO needs to be informed of the model parameters and operational constraints of the HVAC systems and building thermal zones, which the CBOs and end-users are either reluctant to provide or simply unaware of. Therefore, it is practically challenging to implement the bi-level structure for optimal pricing and DR scheduling.

## 2. Framework of the Proposed Strategy for Optimal Retail Pricing Considering Optimal DR of HVAC Units

### 2.1. Original contributions

This paper proposes a new strategy for optimal retail electricity pricing considering the optimal price-based DR of HVAC systems, based on the ANN-based, single-level decision-making structure, rather than the physics-model- based, bi-level structure. The proposed strategy is simple and successfully reflects game-theoretic relations between the DSO and CBOs using online prediction of the optimal data-driven DR schedules for the day-ahead retail price.

In the proposed strategy, online supervised learning is used extensively to develop ANN-based data-driven models of the building thermal dynamics and facilitate applications to the optimal DR. Specifically, the initial training of an ANN is performed using historical data on building operations. In general, the historical data vary little, because for the non-DR case, the operating profiles of the HVAC units do not change significantly unless the occupants complain. The ANN then undergoes repeated online training, as the optimal DR is initiated and the new data on the HVAC system operations and indoor temperature variations start to be obtained for various profiles of electricity price and building thermal environment. The online training is also carried out to learn the monthly variations in weather conditions. This enables the ANN to adapt well to changes in the building operating conditions, gradually mitigating over-fitting and improving model accuracy.

To determine the optimal DR of the HVAC system, an optimization problem is formulated using an ANN-based thermal dynamic model. In particular, the ANN-based model is represented as a "white box" in the proposed strategy, where an explicit set of linear and nonlinear equations describes the complete architecture of the ANN. The equation set can be directly integrated into the optimization problem for the price-based DR. An off-the-shelf deterministic solver is then readily applicable to the problem, ensuring near-global optimality and high reliability of the solution within a reasonable computational time. This approach distinguishes this paper from previous studies, where the ANN-based model was treated as a black box (e.g., [19] and [20]). The black-box model requires a heuristic solver, such as genetic and firefly algorithms, to determine the optimal DR schedule. At every run, the heuristic solver is likely to return different solutions due to the use of random variables, most of which fall into local optima. Therefore, it is necessary to solve the same problem several times, eventually increasing the computation time and making it difficult for the ANN-based DR to be implemented practically.

Furthermore, in the proposed strategy, the DR optimization problem is solved offline for each historical electricity price and building environment dataset. As the optimal solutions are collected, another ANN can be implemented and trained to directly predict the optimal DR schedules for the day-ahead electricity prices, given the forecasts of the outdoor and indoor thermal environments of the building; this is referred to as meta-prediction (MP) in this paper. The MP- enabled ANN reflects the characteristics of the optimal DR of the HVAC system for the time-varying electricity price, where the thermal dynamics of the ANN-based building model is also incorporated. In other words, the MP-enabled ANN can replace the DR optimization problem. The optimal pricing strategy can then be implemented in a single-level decision-making structure via integration with the MP-enabled ANN, unlike the conventional, bi-level one that results from the integration with the DR optimization problem, as shown in Fig. 1. Therefore, the proposed pricing strategy is simpler and more practical.

White-box modeling is also used to integrate MP-enabled ANNs, as in the case of the optimal DR strategy. In other words, an explicit equation set is obtained to describe the architecture of the MP-enabled ANN, and used as the constraint set in the pricing optimization problem for the application of a deterministic solver. Furthermore, the MP-enabled ANN is updated via online learning, as new data on electricity prices and power inputs of the HVAC systems are obtained for various conditions of the distribution grid operation and building thermal environment. This facilitates the practical implementation of the proposed pricing strategy, particularly in the case such that the data on the optimal DR schedules are not sufficient in the early stages. Moreover, the CBOs or end-users can provide the DSO with parameter sets for MP-enabled ANNs, which are more easily obtainable and less privacy sensitive than the physics-based model parameters and operational constraints of the HVAC systems and building thermal zones. To the best of our knowledge, this is the first study to propose the optimal electricity pricing using the online MP of the optimal price-based DR of the HVAC units, to relieve the need for bi-level decision-making structures while still reflecting the hierarchical game-theoretic relations between the DSO and CBOs.

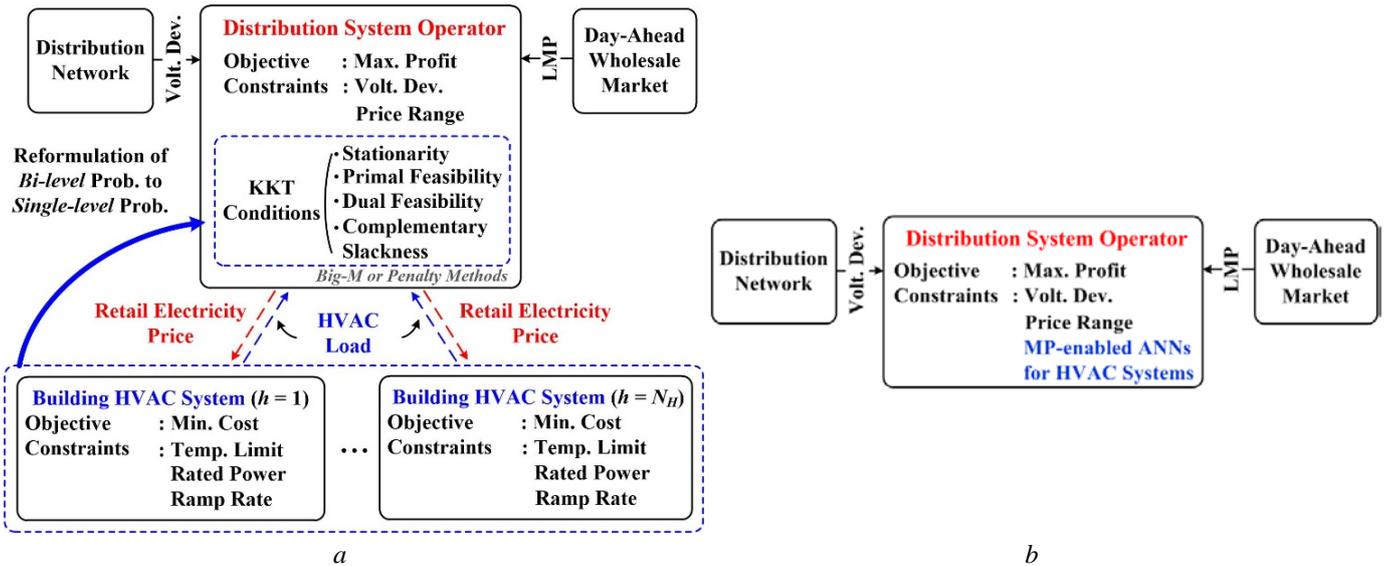

**Fig. 1**. *Comparisons of the conventional and proposed pricing strategies considering the optimal DR of the HVAC systems in commercial buildings*
*a* Conventional pricing
*b* Proposed pricing

The contributions of this paper are summarized as follows:
• An MP-enabled ANN is developed to reflect the characteristics of the optimal DR of the HVAC unit for the electricity price and building thermal environment. Using the MP-enabled ANN, the optimal pricing strategy of the DSO can be established in the single-level decision-making structure, which is simpler and more practical than the conventional, bi-level one.

• Online supervised learning is applied to both the ANN- based building model and the MP-enabled ANN model. This facilitates initial implementation of the data-driven DR and retail pricing strategies, gradually improving the model accuracy and scheduling performance as the size and variability of the training data continue to increase.

• The architectures of the MP-enabled ANN and the ANN- based building model are described using explicit linear and non-linear equations. These can be directly integrated into the optimization problems for the electricity pricing and price- based DR, respectively. A deterministic solver can then be applied, ensuring the near-global optimality and high reliability of the solutions within a reasonable computational time.

• The proposed single-level strategy is tested for a large-scale distribution network including a number of HVAC systems. The case study results verify that the strategy can successfully reflect the complex, hierarchical game-theoretic relations between the DSO and CBOs, while ensuring the occupants' thermal comfort and voltage stability in the distribution grid.

## 2.2. Data flows and algorithm modules in the proposed strategy

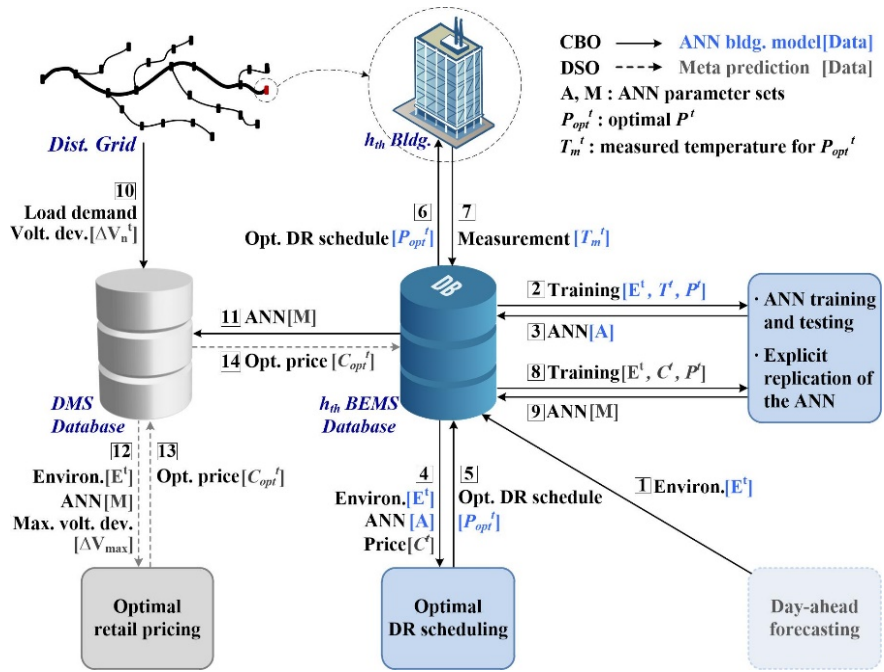

**Fig. 2**. Overall schematic diagram (i.e., data flows and algorithm modules) of the proposed pricing strategy considering the optimal DR of the HVAC units

Fig. 2 shows an overall schematic diagram of the proposed pricing strategy considering the optimal DR of the HVAC systems, focusing on the data flows between the algorithm modules through the databases of the BEMS and distribution management system (DMS). The arrows numbered 1 to 7 represent the data flows for the optimal DR strategy. Specifically, for each building, an ANN is trained using historical data (i.e., [$\mathbf{E^t}$, $T^t$, $P^t$]) to develop a data-driven model of the building thermal dynamics (or equivalently to estimate $T^t$ for $P^t$ and $\mathbf{E^t}$). In the training data, $\mathbf{E^t}$ represents the outdoor and indoor thermal environments of the building. For the day-ahead DR scheduling, $\mathbf{E^t}$ needs to be forecasted; for brevity, forecasting algorithms are not discussed in this paper. Moreover, $T^t$ and $P^t$ are the indoor temperature and HVAC load, respectively. Once the ANN is trained, it is replicated using an explicit set of linear and nonlinear equations with parameter set $\mathbf{A}$, which completely describes the ANN architecture: e.g., number of layers, types of activation functions, and coefficients for neurons. The optimization problem, discussed in Section III, can then be formulated for the price-based DR of the HVAC system using the equation set with $\mathbf{A}$, the forecasted $\mathbf{E^t}$, and the electricity price $C^t$. The BEMS delivers the optimal solution $P_{opt}^t$ to the VSD of the HVAC unit as the reference power input via the communications links, and then receives the measured data on the corresponding $T_m^t$. Note that we adopt a direct control method for the HVAC system [2], where the reference power input $P_{ref}^t$ is directly adjusted to maintain $T^t$ within an acceptable range. As $P_{ref}^t$ is almost the same as $P^t$ due to the fast time response of the VSD, $P_{ref}^t$ does not need to be included in the training data.

In Fig. 2, the data flows corresponding to numbers 2 and 3 are active particularly when the ANN-based, data-driven model is continuously trained via online learning. In general, the performance of the ANN-based model improves gradually as the size and variability of the data pertaining to the HVAC system operations and building thermal conditions increase. In other words, $T^t$ estimated by the ANN becomes very similar to $T_m^t$ measured in the building.

Furthermore, the arrows with numbered 8 to 14 indicate the data flows for the proposed pricing strategy using the MP-enabled ANN. More specifically, an ANN is trained offline with historical data on the inputs and output of the optimal DR strategy (i.e., [$\mathbf{E^t}$, $C^t$, $P^t$]), so that it reflects the characteristics of the optimal response of the HVAC system to the time-varying electricity price and building environment (or more accurately, to predict the optimal $P^t$ for $C^t$ and $\mathbf{E^t}$). As in the case of the ANN-based building model, the trained ANN is represented using explicit equations with parameter set $\mathbf{M}$. The CBO or end-users can provide $\mathbf{M}$ to the DSO as a condition of participating in the price-based DR program, as they submit the operational parameters and constraints of the HVAC systems and building thermal zones in conventional strategies [12], [18]. Therefore, no significant modification is required to the DMS and BEMSs. The optimization problem, discussed in Section IV, is formulated using the forecasted $\mathbf{E}^t$ and the ANN equations with $\mathbf{M}$, considering the voltage deviations $\Delta \mathbf{V_n^t}$ in the distribution grid. The DSO announces the day-ahead profile of $C_{opt}^t$ and the CBOs then perform the optimal DR for $C_{opt}^t$.

Analogously, the data flows specified by the arrows numbered 8 and 9 are active particularly during the period when the online learning is applied to the MP-enabled ANN. The performance of the MP-enabled ANN, and consequently the proposed pricing strategy, improves gradually, as the training data [$\mathbf{E}^t$, $C^t$, $P^t$] are updated with the new profiles of $C_{opt}^t$ and $P_{opt}^t$ obtained for the daily operating conditions of the distribution grid and building thermal environment. In other words, given $C_{opt}^t$, the profiles of $P_{opt}^t$ predicted using the MP-enabled ANN and corresponding $T^t$ estimated using the ANN-based building model become similar to those obtained by solving the DR optimization problem.

## 3. Optimal DR of an HVAC Unit Using the ANN-Based, Data-Driven Model of Building Thermal Dynamics

*3.1. ANN-based model of building thermal dynamics*

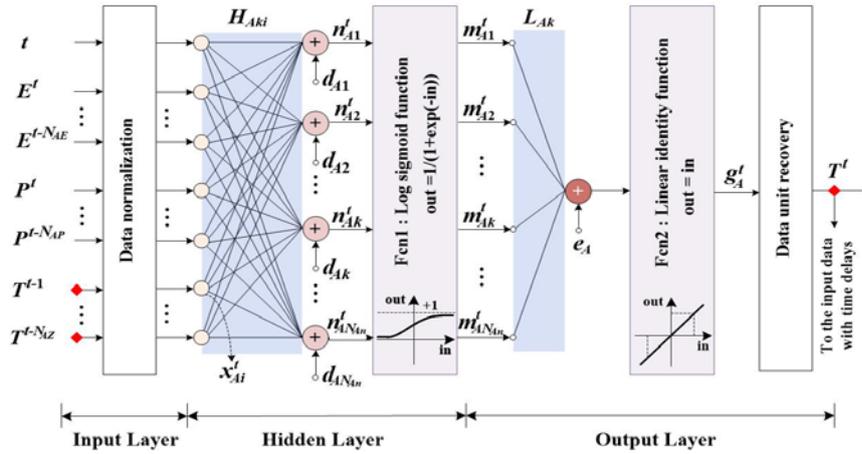

*Fig. 3*. ANN-based model to estimate the indoor temperature for the power input of the HVAC unit, given the building thermal environmental conditions

An ANN-based data-driven model is implemented to reflect the thermal dynamics of a single-zone building: i.e., the variation in $T^t$ for the change in $P^t$, given $\mathbf{E}^t$. In general, $T^t$ at the current time is affected by the past values of $P^t$ and $\mathbf{E}^t$, because of the large thermal capacity inherent in the building structure. Moreover, for the same values of $P^t$ and $\mathbf{E}^t$, $T^t$ varies based on its previous values. In other words, $T^t$ serves as both the state and output variables of the building thermal dynamic model. This implies that the thermal dynamics can be represented successfully using the ANN model with the time-delayed inputs of $P^t$ and $\mathbf{E}^t$, and the feedback loops of $T^t$. For example, the ANN can be designed in the form of a nonlinear auto-regressive network with exogenous inputs (NARX), as shown in Fig. 3. In this paper, the numbers of time-delayed and feedback inputs $N_{AE}$, $N_{AP}$, and $N_{AZ}$ were determined using a trial-and-error approach, considering the trade-off between the modelling accuracy of the ANN and the complexity of the optimization problem, as discussed in Section III-B. For the same reason, a NARX was chosen with one hidden layer. Note that the application of this strategy to a multi-zone building will require an increase the depth of the hidden layers and hence impose additional constraints on the relationship between consecutive hidden layers in the optimization problem. The training and test performances of the ANN were evaluated using the normalized mean squared error (NMSE) as:

$$e(T) = 1 - \frac{\sum_{t}^{N_T}\left(T^t - T^{t\prime}\right)^2}{\sum_{t}^{N_T}\left(T^t - T_{avg}\right)^2}, \quad (1)$$

where $T^{t\prime}$ is the predicted value of $T^t$ and $T_{avg}$ is the average of $T^t$ in the training or test dataset. Moreover, $N_T$ is the number of training or test data points.

Overfitting is a common issue in supervised learning. This occurs when an ANN is fitted to a particular set of training data and fails to reflect the general characteristics of the target model. Numerous methods have been proposed to mitigate overfitting [21], [22], [23], such as weight decay, random dropout, and early stopping of training, which commonly aim to select the least over-fitted network of the candidates. This paper adopts the approach, discussed in [23], where the overfitting index is evaluated to select an appropriate ANN architecture. Briefly, the index is estimated as low, when for constant $\mathbf{E}^t$, $T^t$ increases gradually as $P^t$ decreases steadily to zero and, at the same time, $T^t$ decreases as $P^t$ increases to $P_{max}$. The overfitting index is calculated repeatedly while the network architecture is varied within a pre-determined search range defined by, for example, the maximum time delays of the input and feedback data and the numbers of the activation function types and hidden neurons and layers. We select the ANN that has the lowest overfitting index and an NMSE higher than a pre-specified threshold. We then integrate the ANN into the DR optimization problem, discussed in Section III-B, and apply online learning to the ANN.

This integration is readily achieved, because the connections between the input and output nodes of each layer, shown in Fig. 3, can be represented using linear equations, apart from the nodes for the activation function $Fcn_1$. Specifically, the $i_{th}$ input neuron $x_{Ai}^t$ and the $k_{th}$ hidden neuron $n_{Ak}^t$ are linearly related as:

$$n_{Ak}^t = \sum_{i=1}^{N_{AF}} H_{Aki} x_{Ai}^t + d_{Ak}, \quad \forall i, \forall k, \forall t, \quad (2)$$

where $H_{Aki}$ is the weighting coefficient and $d_{Ak}$ is the bias value. Note that $x_{Ai}^t$ is the normalized value, ranging from –1 to 1, of the $i_{th}$ training data $X_{Ai}^t$, given as:

$$-\frac{2}{(X_{Ai,\max} - X_{Ai,\min})} X_{Ai}^t + x_{Ai}^t = -\frac{2 X_{Ai,\min}}{(X_{Ai,\max} - X_{Ai,\min})} - 1, \quad \forall i, \forall t, \tag{3}$$

where $X_{Ai,max}$ and $X_{Ai,min}$ are the maximum and minimum values, respectively, of $X_{Ai}^t$ in the training dataset. Note that for all $i$, $X_{Ai}^t$ is finite and time-varying. Furthermore, for $n_{Ak}^t$ in (2), the output node of $Fcn_1$ is estimated as:

$$m_{Ak}^t = Fcn_1(n_{Ak}^t) = \{1 + \exp(-n_{Ak}^t)\}^{-1}, \quad \forall k, \forall t, \tag{4}$$

where a log sigmoid function is adopted; this function has been commonly used for general types of ANN [24], [25]. Moreover, as in (2), the output neuron $g_A^t$ is related to $m_{Ak}^t$ in (4) as:

$$g_A^t = \sum_{k=1}^{N_{An}} L_{Ak} \cdot m_{Ak}^t + e_A, \quad \forall k, \forall t, \tag{5}$$

where $L_{Ak}$ is the weighting coefficient and $e_A$ is the bias value. As $g_A^t$ is the normalized value of $T^t$ due to the pre-processor (3); the post-processor is required to reverse transform $g_A^t$ back into the same unit as the original training data $T^t$ as:

$$-\frac{(T_{Ao,\max} - T_{Ao,\min})}{2} g_A^t + T^t = \frac{(T_{Ao,\max} - T_{Ao,\min})}{2} + T_{Ao,\min}, \quad \forall t. \tag{6}$$

where $T_{Ao,max}$ and $T_{Ao,min}$ are the maximum and minimum temperatures, respectively, in the training dataset. Note that the parameter set $\mathbf{A} = [H_{Aki}, L_{Ak}, d_{Ak}, e_A, T_{Ao,max}, T_{Ao,min}, X_{Ai,max}, X_{Ai,min}]$ in (2)–(6) can be easily extracted from the trained ANN and is continuously updated when online learning is applied. Without loss of generality, (2)–(6) can be used with different architectures and activation functions of the ANN.

*3.2. Optimization problem formulation for price-based DR*

Using the ANN-based thermal dynamic model (i.e., (2)–(6)), the optimal DR of the HVAC system is achieved by solving

$$\arg\min_{P^t} J_C = \sum_{t=1}^{N_T} C^t \cdot P^t, \tag{7}$$

subject to the following:

- *Constraints on the physical conditions on $T^t$ and $P^t$*

$$T_{\min}^t \leq T^t \leq T_{\max}^t, \quad \forall t, \tag{8}$$

$$0 \leq P^t \leq P_{\max}, \quad \forall t, \tag{9}$$

$$R_N \leq \left(P^t - P^{t-\Delta t}\right)/\Delta t \leq R_P, \quad \forall t, \tag{10}$$

- *Constraints on the relations among $T^t$, $P^t$, and $\mathbf{E}^t$ of the ANN*

$$n_{Ak}^t - \sum_{i \in \mathbf{X}_{AP}}^{N_{AP}+1} H_{Aki} x_{Ai}^t - \sum_{i \in \mathbf{X}_{AZ}}^{N_{AZ}} H_{Aki} x_{Ai}^t = \sum_{i \in \mathbf{X}_{AE}}^{N_{AE}+2} H_{Aki} x_{Ai}^t + d_{Ak}, \tag{11}$$
$$\forall i, \forall k, \forall t,$$

and (3)–(6),

- *Constraints on the current and time-delayed inputs of the ANN*

$$\boldsymbol{\alpha}_{\mathbf{Aj}}^{\mathbf{t}} = \mathbf{E}^{\mathbf{t}\text{-}(\mathbf{j}\text{-}2)\Delta \mathbf{t}}, \quad \forall j \geq 2,\ t \geq j\Delta t, \tag{12}$$

$$\boldsymbol{\alpha}_{\mathbf{Aj}}^{\mathbf{t}} = \mathbf{E}_{\mathbf{pre}}^{(\mathbf{N_T}+(\mathbf{t}\text{-}(\mathbf{j}\text{-}2)\Delta \mathbf{t}))}, \quad \forall j \geq 2,\ t \leq (j-1)\Delta t, \tag{13}$$

$$\beta_{Al}^t = P^{t-(l-1)\Delta t}, \quad \forall l,\ t \geq l\Delta t, \tag{14}$$

$$\beta_{Al}^t = P_{pre}^{(N_T+(t-(l-1)\Delta t))}, \quad \forall l,\ t \leq (l-1)\Delta t, \tag{15}$$

$$\gamma_{Af}^t = T^{t-f\Delta t}, \quad \forall f,\ t \geq (f+1)\Delta t, \tag{16}$$

$$\gamma_{Af}^t = T_{pre}^{(N_T+(t-f\Delta t))}, \quad \forall f,\ t \leq f\Delta t, \tag{17}$$

$$X_{Ai}^t \in \left\{ t,\ \boldsymbol{\alpha}_{\mathbf{Aj}=2}^{\mathbf{t}},\ \cdots,\ \boldsymbol{\alpha}_{\mathbf{Aj}=(N_{AE}+2)}^{\mathbf{t}},\ \beta_{Al=1}^t,\ \cdots,\ \beta_{Al=(N_{AP}+1)}^t,\right.$$
$$\left. \gamma_{Af=1}^t,\ \cdots,\ \gamma_{Af=N_Z}^t \right\}, \quad \forall i, \forall t. \tag{18}$$

As shown in the objective function (7), the CBO aims to minimize the daily operating cost $J_C$ of the HVAC system for the time-varying $C^t$. In the first set of the constraints, (8) shows that $T^t$ should be maintained within an acceptable range to ensure the occupants' thermal comfort. Moreover, (9) requires $P^t$ to be less than its rated value $P_{max}$. The constraint (10) specifies the limits on the positive and negative ramp rates (i.e., $R_P$ and $R_N$, respectively) of $P^t$ for the unit time interval $\Delta t = 1$ h.

The second constraint set consists of (2)–(6) to establish the relationship between $P^t$ and $T^t$ for $\mathbf{E^t}$ using the ANN-based model of the building thermal dynamics, rather than the physics-based one. Note that (2) can be equivalently expressed as (11), where $x_{Ai}^t$ is divided into three groups: i.e., the controllable and feedback variables (i.e., $P^t$ and $T^t$ for $i \in \mathbf{X_{AP}}$ and $\mathbf{X_{AZ}}$, respectively) on the left-hand side and the time and environmental parameters (i.e., $t$ and $\mathbf{E^t}$ for $i \in \mathbf{X_{AE}}$) on the right-hand side.

Furthermore, (12)–(17) indicate the constraints on the current and time-delayed values of $\mathbf{E^t}$ and $P^t$, as well as the time-delayed values of $T^t$, which are delivered to the input neurons after the data normalization. Specifically, (13), (15), and (17) represent the fact that the building operating data measured in the late evening of day $d$–1 are required to determine the optimal DR schedule for the early morning of day $d$. The constraints (12), (14), and (16) show that after the early morning, the input neurons receive the forecasted parameters $\mathbf{E^t}$ and the decision variable $P^t$, as well as the corresponding feedback response $T^t$. Moreover, (18) lists the types of data that are fed to the input neurons, connecting the second and third constraint sets.

Due to the explicit expressions of the constraints, an off-the- shelf nonlinear programming (NLP) solver can be applied to (3)–(18), ensuring the near-global optimality of the solution. In addition, the same formulation of (3)–(18) can be used for various types of HVAC units and building thermal zones, with only slight modifications of parameter set $\mathbf{A}$.

## 4. Retail Pricing for HVAC Units to Maximize DSO's Profit Considering Distribution Grid Voltages

### 4.1. Meta-prediction of optimal DR to electricity price

Fig. 4*a* shows a simplified schematic diagram of the proposed electricity pricing strategy, focusing on the data flows for the online training and application of the MP-enabled ANN. Specifically, the optimization problem (3)–(18) is repeatedly solved offline to obtain $P_{opt}^t$ and the corresponding $T^t$ for the historical data $C^t$ and $\mathbf{E^t}$. The dataset of [$\mathbf{E^t}$, $C^t$, $P^t$] can then be established to train another ANN, in which the output is set to $P_{opt}^t$ and the inputs include the current and time-delayed values of [$t$, $\mathbf{E^t}$, $C^t$] and the feedback values of $P_{opt}^t$, as shown in Fig. 4*b*. After training, the ANN becomes capable of directly predicting (or *meta-predicting*) the optimal power inputs of the HVAC unit for the day-ahead electricity

price, given the forecast of the building thermal environment. In other words, the MP-enabled ANN work as a price-and-optimal-demand curve. Using the ANN-based curve, the proposed pricing strategy can be developed in a single-level structure, as discussed in Section IV-B, without the need to solve (3)–(18).

For simplicity, the MP-enabled ANN is also designed in the form of NARX, shown in Fig. 3. As in Section III-A, a specific architecture of the NARX has been selected using the NMSE $e(P)$ and the overfitting index, which are calculated by replacing $T^t$ in (1) with $P^t$ and measuring the variations in $P^t$ for the changes in $C^t$, respectively. The MP-enabled ANN is then transformed into a white-box model with parameter set $\mathbf{M} = [H_{Mki}, L_{Mk}, d_{Mk}, e_M, P_{Mo,max}, P_{Mo,min}, X_{Mi,max}, X_{Mi,min}]$, which can be integrated directly into the optimal pricing strategy. After the pricing strategy has been initiated, the online learning is applied to the MP-enabled ANN, as shown in Fig. 4$a$. In general, this improves the MP performance, as the new profiles of $C_{opt}^t$ and $P_{opt}^t$ are obtained for various conditions on $\mathbf{E^t}$.

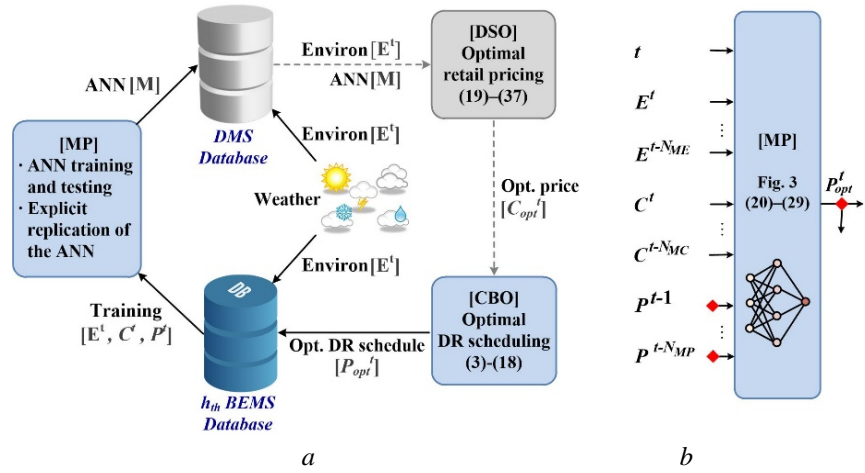

**Fig. 4**. *Simplified diagram of the proposed pricing strategy using the online MP*
*a* Data flows
*b* Inputs and output of the MP-enabled ANN

### 4.2. Optimization problem formulation for retail pricing

Using the MP-enabled ANN, optimal retail pricing is achieved by solving the single-level optimization problem as:

$$\arg\max_{C^t, P^{hvt}} J_D = \sum_{t=1}^{N_T} \sum_{v \in \mathbf{V_B}}^{N_B} \sum_{h=1}^{N_H^v} (C^t - L^t) \cdot P^{hvt}, \quad (19)$$

subject to the following:
• *Constraints on the relations among $C^t$, $P^t$, and $\mathbf{E^t}$ of the ANN*

$$n_{Mk}^{hvt} - \sum_{i \in \mathbf{X_{MC}^{hv}}}^{N_{MC}^{hv}+1} H_{Mki}^{hv} x_{Mi}^{hvt} - \sum_{i \in \mathbf{X_{MP}^{hv}}}^{N_{MP}^{hv}} H_{Mki}^{hv} x_{Mi}^{hvt} = \sum_{i \in \mathbf{X_{ME}^{hv}}}^{N_{ME}^{hv}+2} H_{Mki}^{hv} x_{Mi}^{hvt} + d_{Mk}^{hv}, \tag{20}$$

$$\forall i, \ \forall k, \ \forall t, \ \forall h, \ \forall v,$$

$$(3)\text{--}(6) \text{ using the set of } \mathbf{M}, \tag{21)–(24}$$

- *Constraints on the current and time-delayed inputs of the ANN*

$$\omega_{Mq}^{hvt} = C^{t-(q-1)\Delta t}, \quad \forall q, \ \forall h, \ \forall v, \ t \geq q\Delta t, \tag{25}$$

$$\omega_{Mq}^{hvt} = C_{pre}^{(N_T + (t-(q-1)\Delta t))}, \quad \forall q, \ \forall h, \ \forall v, \ t \leq (q-1)\Delta t, \tag{26}$$

$$\beta_{Ml}^{hvt} = P^{hv(t-l\Delta t)}, \quad \forall l, \ \forall h, \ \forall v, \ t \geq (l+1)\Delta t, \tag{27}$$

$$\beta_{Ml}^{hvt} = P_{pre}^{hv(N_T + (t-l\Delta t))}, \quad \forall l, \ \forall h, \ \forall v, \ t \leq l\Delta t, \tag{28}$$

$$X_{Mi}^{hvt} \in \left\{ t, \boldsymbol{\alpha}_{\mathbf{Mj}=2}^{\mathbf{hvt}}, \cdots, \boldsymbol{\alpha}_{\mathbf{Mj}=(N_{ME}^{hv}+2)}^{\mathbf{hvt}}, \omega_{Mq=1}^{hvt}, \cdots, \omega_{Mq=(N_{MC}^{hv}+1)}^{hvt}, \right.$$
$$\left. \beta_{Ml=1}^{hvt}, \cdots, \beta_{Ml=N_{MP}^{hv}}^{hvt} \right\}, \quad \forall i, \ \forall t, \ \forall h, \ \forall v, \tag{29}$$

and (12) and (13),

- *Constraints on the operational conditions on $C^t$ and $\Delta \mathbf{Vn^t}$*

$$C_{\min}^t \leq C^t \leq C_{\max}^t, \quad \forall t, \tag{30}$$

$$C_N \leq \left(C^t - C^{t-\Delta t}\right)/\Delta t \leq C_P, \quad \forall t, \tag{31}$$

$$-\boldsymbol{\Delta V_{max}} \leq \boldsymbol{\Delta V_n^t} = \begin{bmatrix} \mathbf{S_{V,P}^t} & \mathbf{S_{V,Q}^t} \end{bmatrix} \begin{bmatrix} \boldsymbol{\Delta P^t} & \boldsymbol{\Delta Q^t} \end{bmatrix}^T$$
$$\leq \boldsymbol{\Delta V_{max}}, \quad \text{for } t_{ps} \leq t \leq t_{pe}, \ \forall n, \tag{32}$$

$$\sum_{h=1}^{N_H^v} P^{hvt} + \sum_{h=1}^{N_X^v} P_c^{hvt} = P_o^{vt}, \quad \forall t, \ \forall v, \tag{33}$$

$$\mathbf{P_o^{vt}} = \begin{bmatrix} P_o^{1t}, & P_o^{13t}, & \cdots & P_o^{vt}, & \cdots & P_o^{101t} \end{bmatrix}^T, \quad \forall t, \ \forall v, \tag{34}$$

$$\mathbf{B} = \begin{bmatrix} \mathbf{b^{11}} & \mathbf{b^{12}} & \cdots & \mathbf{b^{1v}} & \cdots \\ \mathbf{b^{21}} & \mathbf{b^{22}} & \cdots & \mathbf{b^{2v}} & \cdots \\ \vdots & \vdots & \ddots & \vdots & \cdots \\ \mathbf{b^{n1}} & \mathbf{b^{n2}} & \cdots & \mathbf{b^{nv}} & \cdots \\ \vdots & \vdots & \vdots & \vdots & \ddots \end{bmatrix}, \quad \forall n, \forall v, \tag{35}$$

$$\mathbf{b^{nv}} = \begin{cases} \left[\begin{matrix} \frac{1}{3} & \frac{1}{3} & \frac{1}{3} & \mathbf{O_{1\times 3}} \end{matrix}\right]^T, & \text{for } n = v, \\ \mathbf{O_{6\times 1}}, & \text{for } n \neq v, \end{cases} \tag{36}$$

$$\begin{bmatrix} \mathbf{\Delta P^t} & \mathbf{\Delta Q^t} \end{bmatrix}^T = \mathbf{B} \cdot \mathbf{P_o^{vt}} = \left[\begin{bmatrix} P_{o,a}^{1t} & P_{o,b}^{1t} & P_{o,c}^{1t} & \mathbf{O_{1\times 3}} \end{bmatrix}, \mathbf{O_{1\times 6}}, \cdots, \begin{bmatrix} P_{o,a}^{13t} & P_{o,b}^{13t} & P_{o,c}^{13t} & \mathbf{O_{1\times 3}} \end{bmatrix}, \mathbf{O_{1\times 6}}, \right.$$
$$\left. \cdots, \begin{bmatrix} P_{o,a}^{vt} & P_{o,b}^{vt} & P_{o,c}^{vt} & \mathbf{O_{1\times 3}} \end{bmatrix}, \mathbf{O_{1\times 6}}, \cdots \right]^T, \tag{37}$$
$$\forall v, \forall t$$

In (19), the DSO determines the optimal $C^t$ to maximize its profit $J_D$, which is defined as the difference between the revenue earned by selling electricity to the HVAC systems at $C^t$ minus the cost of buying the electricity at $L^t$. By solving (19)–(37), the DSO can also estimate the optimal power inputs $P^{hvt}$ of HVAC unit $h$ at bus $v$ in a distribution grid, induced by $C^t$.

As in Section III-B, the first set of (20)–(29) represents the relationship between $C^t$ and $P^t$ of the MP-enabled ANN for each HVAC unit, given the day-ahead forecast of $\mathbf{E^t}$; $\mathbf{M}$ is used here, instead of $\mathbf{A}$. For example, as in (11), (20) specifies the relationship between the $i_{th}$ input neuron $x_{Mi}^{hvt}$ and the hidden neuron $n_{Mk}^{hvt}$, where $x_{Mi}^{hvt}$ is divided into three groups: i.e., $\mathbf{X_{MC}^{hv}}$, $\mathbf{X_{MP}^{hv}}$ and $\mathbf{X_{ME}^{hv}}$. The terms on the left-hand side include the controllable and feedback variables, and the terms on the right-hand side contain the time and environmental parameters. The second set includes the constraints on the current and time-delayed values of $C^t$ and $\mathbf{E^t}$ (i.e., controllable and independent inputs, respectively) and the time-delayed values of $P^t$ (i.e., feedback inputs) that are fed into the input neurons of the MP-enabled ANN.

In addition, (30) and (31) are necessary to achieve low-level price volatility [26]. The price cap $C_{max}^t$ in (30) aims to mitigate the risk of the DSO by excessively increasing its profit, which is unfair to the CBOs. By (31), the ramp rates of $C^t$ during $\Delta t = 1$ h are limited to prevent abrupt changes in $P^{hvt}$ and consequently severe increases in the operating stress and maintenance costs of the HVAC unit. These constraints can be brought into effect via mutual agreement between the DSO and CBOs [26].

While maximizing the profit, the DSO is responsible for stable operation of the distribution network [18]. In this paper, to ensure the grid voltage stability, (32) specifies that the incremental variations in the voltage $\Delta V_n^t$ at all buses $n \in \mathbf{V_A}$, owing to the total HVAC load demand $[\mathbf{\Delta P^t}, \mathbf{\Delta Q^t}]^T$, should be maintained within the limits ±$\Delta V_{max}$, particularly during the on-peak hours: i.e., $t_{ps} \leq t \leq t_{pe}$. Specifically, $\mathbf{\Delta V_n^t}$ is calculated using the sensitivity matrices $\mathbf{S_{V,P}^t}$ and $\mathbf{S_{V,Q}^t}$, which can be derived from general power flow equations for a three-phase (3-*ph*) distribution grid. In (33), $P_o^{vt}$ is represented as the total sum of the power inputs $P^{hvt}$ and $P_c^{hvt}$ of the price-sensitive and conventional HVAC units, respectively, at bus $v \in \mathbf{V_B}$. A column vector $\mathbf{P_o^{vt}}$ is then established using $P_o^{vt}$ as elements, as shown in (34). Moreover, a conversion matrix $\mathbf{B}$ is defined in (35) and (36), so that $\mathbf{P_o^{vt}}$ is converted into another column vector $[\mathbf{\Delta P^t}, \mathbf{\Delta Q^t}]^T$, as shown in (37), where each element in $\mathbf{P_o^{vt}}$ is relocated considering the topology of the distribution grid. As an example, (37) describes $[\mathbf{\Delta P^t}, \mathbf{\Delta Q^t}]^T$ for the IEEE 123-Node Test Feeder, discussed in Section V, where the price-sensitive and conventional units are connected to the pre-specified buses: i.e., $v \in \mathbf{V_B}$ = {1, 13, 18, 42, 47, 52, 57, 60, 63, 67, 76, 81, 89, 97, 101}. As shown in (37), $[\mathbf{\Delta P^t}, \mathbf{\Delta Q^t}]^T$ contains 6·$N_A$ elements for an $N_A$-bus grid. For $n = v$, each set of six elements represents the 3-*ph* active and reactive power for phases *a*, *b*, and *c* (i.e., $[P_{o,a}^{vt}, P_{o,b}^{vt}, P_{o,c}^{vt}, Q_{o,a}^{vt}, Q_{o,b}^{vt}, Q_{o,c}^{vt}]^T$) of the total HVAC load. For simplicity, the HVAC units are assumed to operate as 3-*ph* balanced loads with a unity power factor: i.e., $P_{o,a}^{vt} = P_{o,b}^{vt} = P_{o,c}^{vt} = P_o^{vt}/3$ and $Q_{o,a}^{vt} = Q_{o,b}^{vt} = Q_{o,c}^{vt} = 0$.

As in (3)–(18), an off-the-shelf NLP solver is applicable to (19)–(37), due to the explicit replication of the MP-enabled ANN. Moreover, the proposed pricing strategy can be readily applied to a distribution grid, including a number of HVAC systems with different DR characteristics, by taking advantage of the MP: i.e., easy determination and access to $\mathbf{M}$.

## 5. Case Studies and Results

*5.1. Test system and simulation conditions*

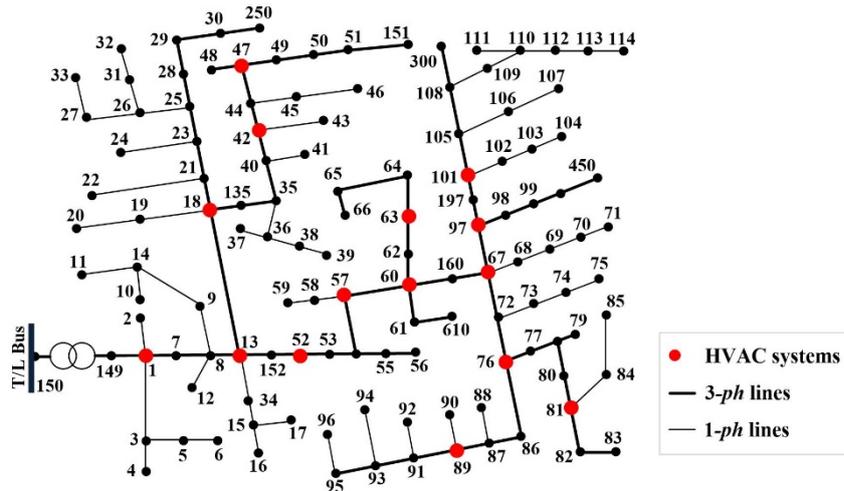

***Fig. 5***. *Test distribution grid for the case studies on the proposed optimal retail pricing strategy using the online MP of the data-driven DR of the HVAC units*

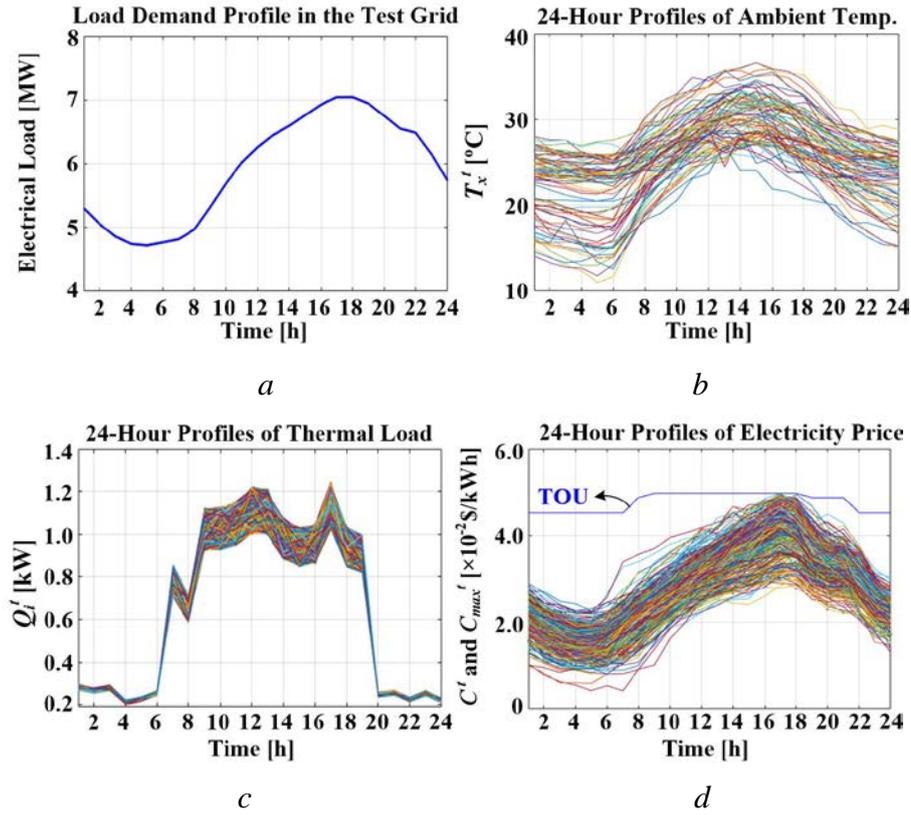

*Fig. 6. Case study conditions*

*a* Total load demand in the test network during a scheduling day
*b* 24-hour profiles of historical ambient temperatures
*c* 24-hour profiles of historical internal thermal loads
*d* 24-hour profiles of historical electricity prices

The proposed strategy was tested using the distribution grid, shown in Fig. 5, which was modeled using the IEEE 123-Node Test Feeder with modifications based on [27]. This includes 1-*ph* loads and 3-*ph* balanced and unbalanced loads with a rated line voltage of 4.16 kV. A 24-hour profile of the load demand during summer time was obtained from [28] and scaled down, so that the peak load demand in the test grid was equal to 7.04 MW, as shown in Fig. 6*a*. For simplicity, the constant-power load model was used for the case studies; the ZIP- coefficient model can also be adopted for the proposed strategy to reflect the steady-state load behaviors more comprehensively.

In addition, commercial buildings containing price-sensitive or conventional HVAC units are located at 15 buses, each of which is marked by a red circle in Fig. 5. The total number and capacity of the HVAC units were set to 12 and 195.8 kW, respectively, at each of the 15 buses. An experimental test building [8] was considered, as shown in Fig. 7, for the data acquisition and comparative case studies. Briefly, the experimental building is divided into test and climate chambers, both of which are within a larger laboratory room in which the indoor temperature is maintained at an almost constant level. The test chamber includes lights and heat sources to simulate the internal thermal loads of a common office, and the climate one contains a separate HVAC system to emulate the outdoor thermal environment. In [8] and [9], a building simulator was developed using an inverse transfer function to estimate $T^t$ for $P^t$, given $\mathbf{E^t}$ = [$T_a^t$, $T_x^t$, $Q_i^t$], where $T_a^t$ and $T_x^t$ are the adjacent and ambient temperatures, respectively, and $Q_i^t$ is the internal thermal load. The building simulator was run using the data of $T_x^t$ measured from April 1 to September 30 in the years 2012 and 2013 in Boston, MA [29], and of $Q_i^t$ estimated based on the operation of a real commercial building [6], [9]; see Fig. 6b and c, respectively. In Fig. 6c, $Q_i^t$ is increased at $t = 7$ h and maintained at high levels until $t = 19$ h, when people start arriving at work and leaving the building, respectively. The size of the initial training data [$t$, $\mathbf{E^t}$, $T^t$, $P^t$] was set to 1,200 (i.e., 50 days) and 16 with respect to time and objects, respectively. The ANN-based thermal dynamic model was trained repeatedly via online learning, as the new profiles of [$\mathbf{E^t}$, $T^t$, $P^t$] that were obtained from (3)–(18).

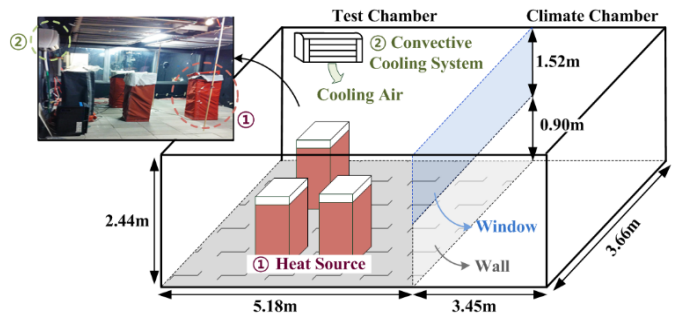

**Fig. 7.** Experimental test building for the data-driven DR of the HVAC system

Furthermore, Fig. 6d shows the 24-hour profiles of $C^t$ at the distribution level [30]. The hourly-varying retail prices can assist the DSO in better exploiting the operational flexibility of DR resources and attracting more consumers to price-based DR than time-of-use (TOU) pricing rates, in which only a few (named on-peak, off-peak, and mid-peak) rates exist and do not change for months [31]. As TOU pricing is simpler and less volatile, we set $C_{max}^t$ in (30) using the profile of the TOU rates [32]. Note that $C_{min}^t$ was set to the wholesale price $L^t$ [30], to guarantee at least the minimum profit of the DSO in the DR business. The MP-enabled ANN was initially trained with the 1,200 datasets of [$t$, $\mathbf{E^t}$, $C^t$, $P^t$] and updated via online learning, as the new profiles of $C_{opt}^t$ were obtained from (19)–(37).

*5.2. Online supervised learning of the ANN-based building model and MP-enabled ANN model*

Fig. 8a shows the online learning results of the ANN-based building model over 190 days. For each day, $e(T)$ was calculated, as shown in Fig. 9a, by comparing the profiles of $T^t$ that were obtained from (3)–(18) and the building simulator, discussed in Section V-A, for the optimal $P^t$. In general, $e(T)$ increased gradually, indicating improvement in the modeling accuracy, as the CBO continued the online learning after initiating the DR strategy. This verifies that the ANN-based, data-driven model can successfully reflect the complex thermal dynamics of the experimental building. After $e(T)$ reached a value higher than 0.99 for five consecutive days, the model started being used to develop the MP-enabled ANN and hence the optimal pricing strategy (i.e., (19)–(37)).

Analogously, Fig. 8*b* shows $e(P)$ of the MP-enabled ANN model. The online learning was performed daily for a further 190 days. To calculate $e(P)$, the 24-hour profiles of $P^t$ were obtained from (19)–(37) using the MP-enabled ANN were compared to those acquired from (3)–(18) for the optimal $C^t$, as shown in Fig. 9*b*. Moreover, the ANN-based building model was trained online with the new building operating data obtained over a period of 190 days. Considering the computational burden, the training was conducted periodically. In Fig. 8*b*, the red arrows indicate periods ranging from approximately 30 days to 45 days. This improved the online learning performance of the MP-enabled ANN further. Noticeable reductions in $e(P)$ were observed occasionally, unlike in the case of $e(T)$; however, it still can be seen that $e(P)$ increased gradually with a gradient of $2.2 \times 10^{-3}$ per day, as the online learning continued. This demonstrates the possibility that the online-trained, MP-enabled ANNs can better reflect the optimal DR characteristics of HVAC systems for time-varying retail prices, as larger volumes of data on the weather, retail price, and building operation are collected with higher granularity. Since there have been few attempts to develop MP-enabled ANN models, the application of more sophisticated ML algorithms is also well worth investigating.

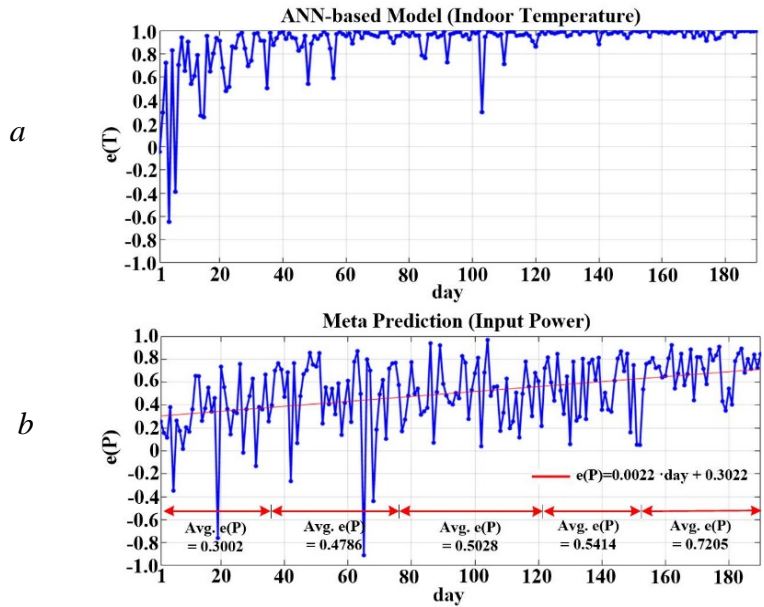

***Fig. 8.*** *Online supervised learning results*
*a* $e(T)$ for the tests of the ANN-based building model
*b* $e(P)$ for the tests of the MP-enabled ANN model

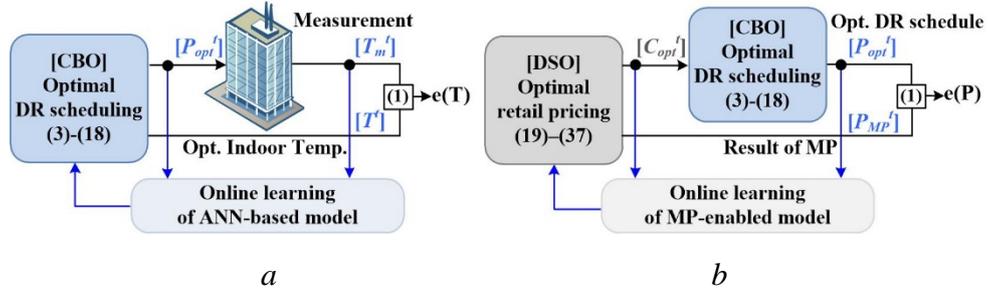

*a*            *b*

**Fig. 9.** *Estimations of e(T) and e(P) for the online supervised learning tests.*

*5.3. Price-based DR using the ANN-based building model*

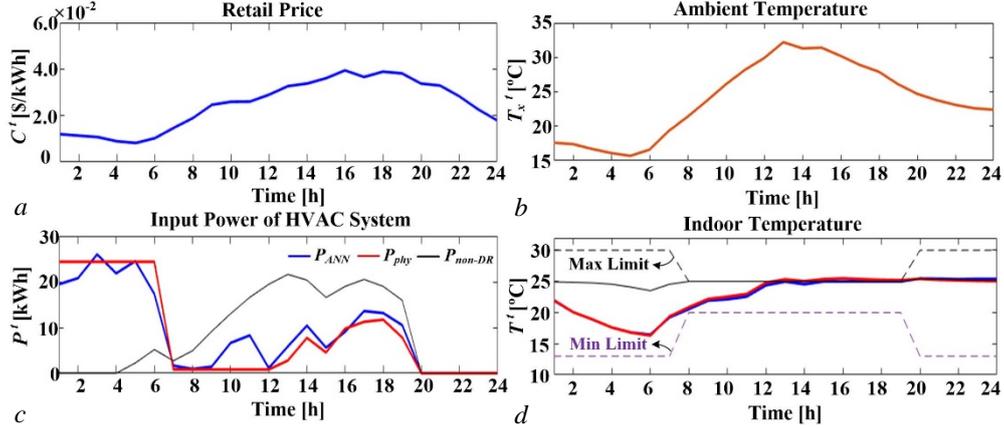

*Fig. 10.* *Optimal price-based DR schedules for the ANN-based model and the fully informed physics-based model*

*a* retail price $C^t$
*b* ambient temperature $T_x^t$
*c* optimal power input $P^t$ of the HVAC system
*d* corresponding indoor temperature $T^t$

**Table 1** Comparisons of the Operating Costs $J_C$ of the HVAC System

|  | Non-DR | Price-based DR | |
| --- | --- | --- | --- |
|  |  | ANN model | Fully informed model |
| Operating cost $J_C$ [$] | 6.60 | 4.33 | 3.67 |
| Reduction rate [%] | - | 34.40 | 44.39 |

Using the ANN-based building model, (3)–(18) was solved to determine the optimal DR schedule of the HVAC system for the 24-hour profiles of $C^t$ and $T_x^t$, shown in Fig. 10*a* and *b*, respectively. Specifically, Fig. 10*c* shows the optimal profile of $P^t$ (i.e., the blue line) for the proposed DR strategy in comparison to that (i.e., the red line) of the ideal case, where the fully informed physics-based model of the building thermal dynamics was used. The profiles were similar to each other, resulting in the comparable operating costs for the HVAC systems. As shown in Table 1, $J_C$ for the proposed case was only $0.66 higher than $J_C$ for the ideal case, and $2.27 lower than $J_C$ for the non-DR case, in which $P^t$ was controlled to maintain $T^t$ at 25°C during 8 h ≤ $t$ ≤ 19 h (see Fig. 10$c$ and $d$). In addition, Fig. 10$d$ represents the corresponding profiles of $T^t$ for the proposed and ideal cases. The differences between the $T^t$ profiles were imperceptible during 24 hours. This verifies that the ANN-based model successfully reflected the building thermal dynamics, enabling the proposed DR strategy to ensure occupants' thermal comfort in practice.

In Fig. 10$c$ and $d$, the large values of $P^t$ were scheduled for 1 h ≤ $t$ ≤ 6 h, whereas $T_x^t$ and $Q_i^t$ were high during the daytime. The HVAC system operated in the pre-cooling mode; $P^t$ was shifted to the early morning when $C^t$ remained low, leading to a reduction in $J_C$. In Fig. 10$d$, the pre-cooled $T^t$ gradually increased from 20°C at $t$ = 8 h to 25°C at $t$ = 13 h, due to the limited thermal capacity of the building structure. To maintain $T^t$ below $T_{max}^t$, $P^t$ began to increase at $t$ = 13 h.

### 5.4. Optimal retail pricing using the MP-enabled ANNs

Using the MP-enabled ANNs, (19)–(37) was solved to determine the optimal $C^t$ and estimate the corresponding $P^t$, $T^t$ and $\Delta \mathbf{V_n^t}$. For simplicity, all entries of $\Delta \mathbf{V_{max}}$ in the voltage constraint (32) were assumed to be the same as $\Delta V_{max}$. The largest voltage deviation $\Delta V^t$ was observed at Bus 95 in the test grid. Fig. 11 represents the optimal solution of (19)–(37) for $\Delta V_{max}$ = 0.07 pu. It can be seen in Fig. 11$d$ that (32) is unbounded: i.e., $|\Delta V^t|$ < $\Delta V_{max}$. Since the voltage stability was already ensured, the DSO did not need to induce the HVAC load shift by reducing $C^t$ during the off-peak hours. Instead, the DSO maintained $C^t$ at high levels for 1 h ≤ $t$ ≤ 11 h to maximize the profit $J_D$. Note that the profile of $C^t$ in Fig. 11$a$ is significantly different to that of $C^t$ shown in Fig. 6$d$. This verifies that the optimal solutions acquired using the online supervised learning were not confined by the historical training data: i.e., the overfitting issue was resolved. Fig. 11$b$ shows that $P^t$ was shifted less to the off-peak hours, compared to $P^t$ in Fig. 10$c$, consequently increasing $J_C$ from \$4.33 to \$7.96.

The case study was repeated for $\Delta V_{max}$ = 0.06 pu and 0.05 pu. The constraint (32) became stricter and hence bounded: i.e., $max(|\Delta V^t|)$ = $\Delta V_{max}$ for $t_s$ ≤ $t$ ≤ $t_e$. The DSO needs to exploit the operational flexibility of the HVAC systems (or equivalently shift $P^t$) to ensure the voltage stability during the on-peak hours by reducing $C^t$ during the off-peak hours. In Fig. 12$a$, $C^t$ for $\Delta V_{max}$ = 0.06 pu was determined to be lower during 1 h ≤ $t$ ≤ 11 h than that for $\Delta V_{max}$ = 0.07 pu. For $\Delta V_{max}$ = 0.05 pu, $C^t$ was further reduced, inducing a larger shift of $P^t$ to satisfy (32). In Figs. 11–13, the values of $\sum_{t=1}^{t_s-1} P^t$ was estimated as 72.9 kWh, 75.2 kWh, and 121.6 kWh, respectively.

In Figs. 11$b$–13$b$, the blue, red, and green lines indicate the optimal profiles of $P^t$ for the three cases, respectively: the pricing strategy (19)–(37), the DR strategy (3)–(18) for the optimal $C^t$, and the DR strategy with the fully informed physics-based model for the optimal $C^t$. It can be seen that the optimal profiles of $P^t$ are similar to each other particularly during the daytime. This demonstrates the accuracy and applicability of the MP-enabled ANN model. The differences among the profiles for $t \leq 5$ h were mainly attributed to the fact that the time-delayed $P^t$ in (15) and (28) was zero during this time period, contributing little to the ANN training. As in Section V-C, the differences in $P^t$ marginally affected $T^t$ (see Figs. 11c–13c), indicating that the proposed pricing strategy ensured occupants' thermal comfort.

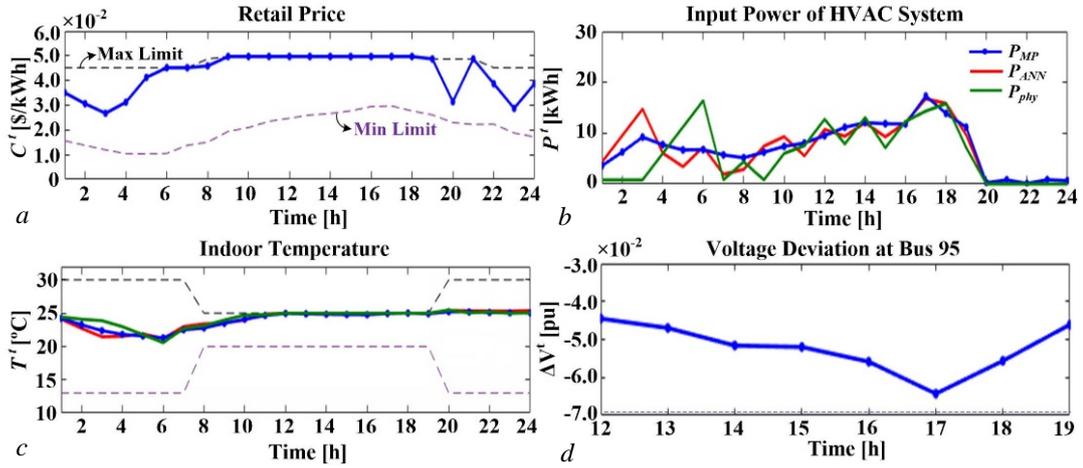

**Fig. 11.** *Optimal results of the proposed retail pricing strategy for $\Delta V_{max} = 0.07$*

a optimal retail price $C^t$
b optimal power input $P^t$ of the HVAC system
c corresponding indoor temperature $T^t$
d corresponding bus voltage variation $\Delta V^t$

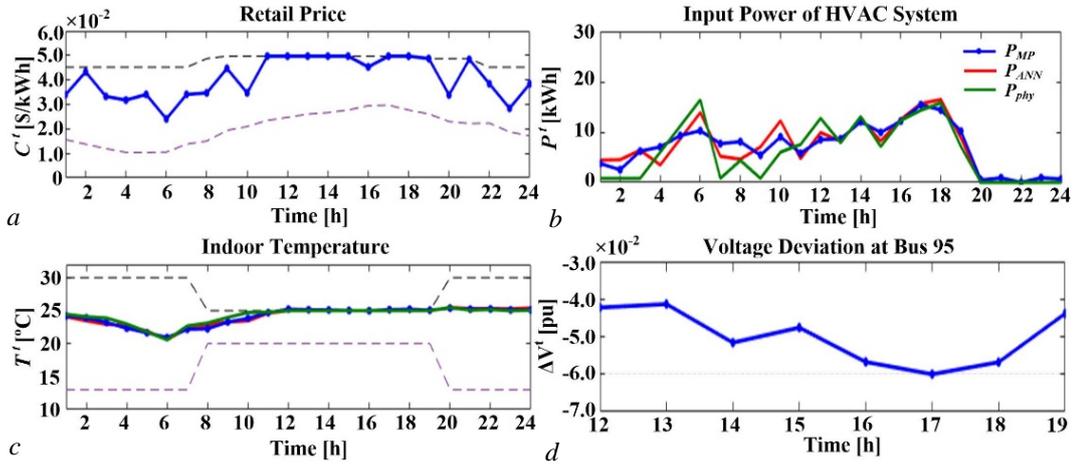

**Fig. 12.** *Optimal results of the proposed retail pricing strategy for $\Delta V_{max} = 0.06$*

a optimal retail price $C^t$
b optimal power input $P^t$ of the HVAC system
c corresponding indoor temperature $T^t$
d corresponding bus voltage variation $\Delta V^t$

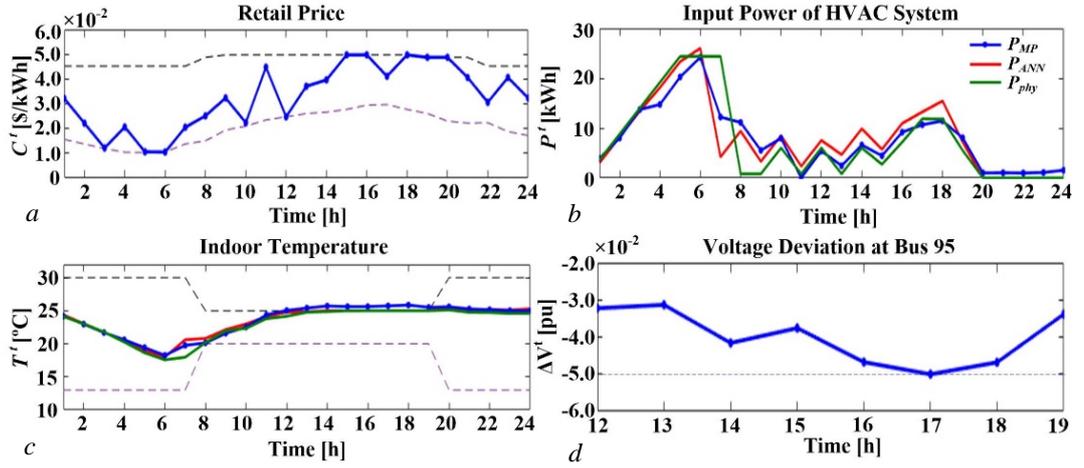

***Fig. 13.*** *Optimal results of the proposed retail pricing strategy for $\Delta V_{max} = 0.05$*

*a* optimal retail price $C^t$
*b* optimal power input $P^t$ of the HVAC system
*c* corresponding indoor temperature $T^t$
*d* corresponding bus voltage variation $\Delta V^t$

**Table 2** Comparisons of the DSO's Profits and CBO's Operating Costs for the Maximum Allowable Deviations of Bus Voltages

| $\Delta V_{max}$ [×10⁻² pu] | DSO's profit, $J_D$ [\$] | | | CBO's operation cost, $J_C$ [\$] | | |
|---|---|---|---|---|---|---|
| | (1) $P_{MP}$ | (2) $P_{ANN}$ | \|(2)–(1)\|/(2) | (1) $P_{MP}$ | (2) $P_{ANN}$ | \|(2)–(1)\|/(2) |
| 7.0 | 430.62 | 405.02 | 0.06 | 836.15 | 795.97 | 0.05 |
| 6.0 | 370.76 | 364.50 | 0.02 | 756.84 | 749.31 | 0.01 |
| 5.0 | 170.03 | 180.66 | 0.06 | 515.02 | 556.34 | 0.07 |

The case study results demonstrated that the proposed pricing strategy enables the DSO to effectively resolve the network voltage issues by adjusting the retail price and inducing the pre-cooling of the HVAC system. As listed in Table 2, this led to a decrease in the DSO's profit; however, it can be sufficiently compensated for by the savings in the installation costs of additional voltage-regulating devices. Note that $J_C$ was also reduced, indicating that the operational flexibility of the HVAC systems became more valuable for smaller values of $\Delta V_{max}$. The results also verified that the proposed simple single-level structure using the MP-enabled ANNs successfully reflected the aforementioned game-theoretic relationships between the DSO and CBOs in the retail electricity market, with consideration of the price-based DR and network voltage stability. Therefore, the DSO can be relieved from the need to use conventional bi-level structures, which are far more complicated and less practical.

## 6. Conclusions

This paper proposed an online ML-based strategy for the optimal retail pricing using the MP of the data-driven DR of the HVAC systems, considering the voltage stability in the distribution grid. The ANN-based model of the building thermal dynamics was developed using online supervised learning and represented with the explicit set of linear and nonlinear equations, requiring no physics-based model parameters. We formulated the DR optimization problem for the HVAC system using the equation set. The optimal $P^t$ and corresponding $T^t$ were obtained offline for various historical conditions of $C^t$ and $\mathbf{E}^t$, resulting in the datasets [$t$, $\mathbf{E}^t$, $C^t$, $P^t$, $T^t$]. Another ANN was then trained online with the datasets for the direct prediction (i.e., MP) of the optimal DR schedules of the HVAC system for the day-ahead time-varying electricity prices. The MP-enabled ANN was used as the price-and-optimal-demand curve, which could replace the DR optimization problem. This enabled the optimal retail pricing to be achieved using the single-level decision-making structure, which is simpler and more practical than the conventional bi-level one. The case studies demonstrated that the proposed strategy successfully reflects the game theoretic relationship between the DSO and CBOs in the retail market with consideration of the optimal price-based DR and distribution grid operation. The DSO and CBOs effectively exploited the operational flexibility of the HVAC units to make the DR application profitable, while ensuring the grid voltage stability and occupants' thermal comfort.